\documentclass[prl,aps,showpacs,twocolumn]{revtex4}
\usepackage{graphicx}

\begin{document}

\title{Unusually strong coherent response from grain-boundary Josephson network in
polycrystalline $Pr_xY_{1-x}Ba_2Cu_3O_{7-\delta}$}

\author{V.A.G. Rivera$^{1}$, S. Sergeenkov$^{1}$, C. Stari$^{1,2}$,
L.Jr. Cichetto$^{1}$, C.A. Cardoso$^{1}$, E. Marega$^{3}$, F.M.
Araujo-Moreira$^{1}$}

\affiliation{$^{1}$Departamento de F\'{i}sica e Engenharia
F\'{i}sica, Grupo de Materiais e Dispositivos, Centro
Multidisciplinar para o Desenvolvimento de Materiais Cer\^amicos,
Universidade Federal de
S\~ao Carlos, S\~ao Carlos, SP, 13565-905 Brazil\\
$^{2}$Instituto de F\'{i}sica, Facultad de Ingenieria, Julio
Herrera y Reissig 565, C.C. 30, 11000, Montevideo, Uruguay\\
$^{3}$ Instituto de F\'{i}sica, USP,  S\~ao Carlos, SP, 13560-970
Brazil}
\date{\today}

\begin{abstract}
By applying a highly sensitive homemade AC susceptibility
technique to $Pr_xY_{1-x}Ba_2Cu_3O_{7-\delta}$ polycrystals (with
$x=0.0$, $0.1$ and $0.3$), we observed very sharp Fraunhofer type
low-field periodic oscillations of the real part of the AC
susceptibility which were attributed to Josephson vortices
penetrating intergranular regions of grain-boundary Josephson
network in our samples. Assuming the Lorentz type distribution of
single-junction contact areas, we were able to successfully fit
the experimental data.
\end{abstract}


\maketitle

\section{Introduction}

It is well-known that important for large-scale applications
properties of any realistic device based on Josephson effects
require a very coherent response from many Josephson contacts
comprising such a device (see, e.g.,~\cite{1,2,3,4,5,6,7,8} and
further references therein). Usually, due to inevitable
distribution of critical currents and sizes of the individual
junctions, a grain-boundary induced Josephson network in
polycrystalline materials manifests itself in a rather incoherent
way, making it virtually impossible for applications. That is why,
ordered artificially prepared (hence more costly) Josephson
junction arrays (JJAs) are used instead to achieve the expected
performance~\cite{9,10,11,12,13,14}.

In this Letter we report on unusually strong coherent response of
grain-boundary Josephson network in our polycrystalline
$Pr_xY_{1-x}Ba_2Cu_3O_{7-\delta}$ (PYBCO) samples which manifest
itself through a clear Fraunhofer type magnetic field dependence
of the measured AC susceptibility (more typical for ordered JJAs).

\begin{figure}
\centerline{\includegraphics[width=8.0cm,angle=0]{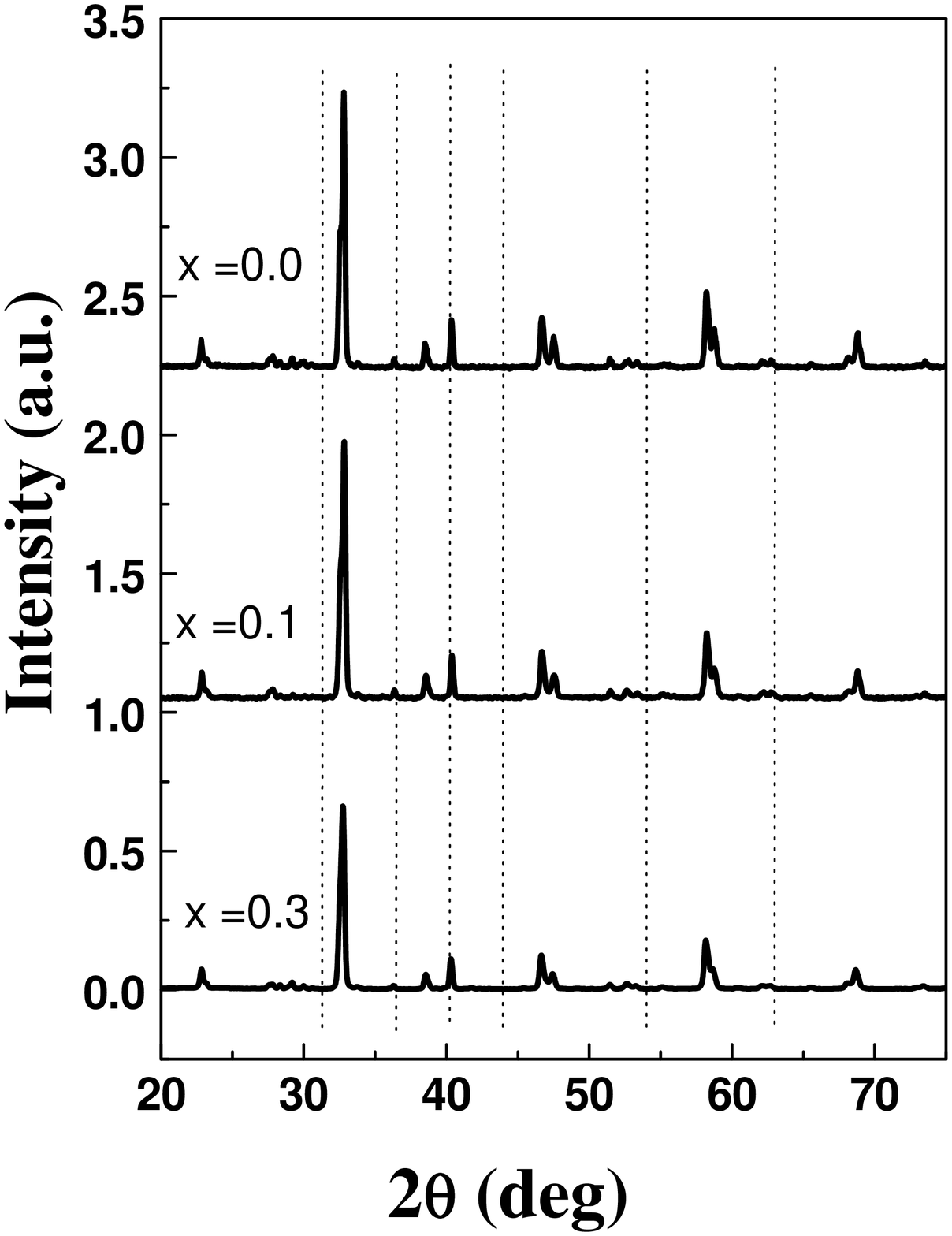}}
\vspace{0.5cm} \caption{XRD patterns of
$Pr_xY_{1-x}Ba_2Cu_3O_{7-\delta}$ samples. }
\end{figure}

\begin{figure}
\centerline{\includegraphics[width=9.0cm,angle=0]{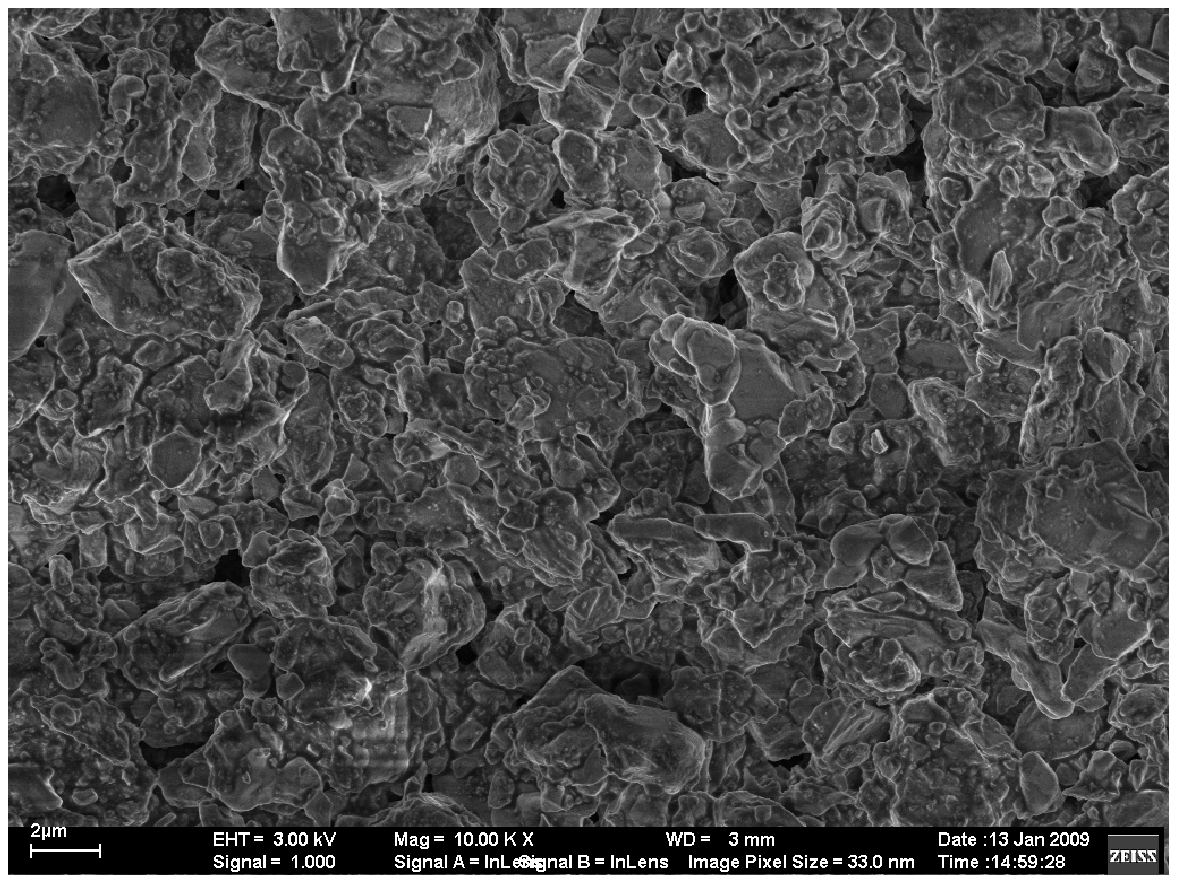}}
\vspace{0.5cm} \caption{SEM scan photography of
$YBa_2Cu_3O_{7-\delta}$. }
\end{figure}

\section{Results}

High quality PYBCO bulk polycrystalline samples have been prepared
by following a chemical route based on the sol-gel
method~\cite{15}. The phase purity and the structural
characteristics of our samples were confirmed by both scanning
electron microscopy (SEM) and x-ray diffraction (XRD) along with
the standard Rietveld analysis. The analysis of the XRD data
(Fig.1) reveals that no secondary phases are present in our
samples and that the peaks correspond to the orthorhombic
structure with $YBa_2Cu_3O_{7-\delta}$ (YBCO) stoichiometric
phase.  The onset temperatures $T_C$ (shown in Fig.3 for all
studied samples) were independently confirmed via the resistivity,
magnetization and AC susceptibility data, and well correlate with
the values reported in the literature for polycrystalline samples
with a similar composition~\cite{16,17}. Fig.2 shows the SEM scan
of grain-boundary morphology in the undoped $YBCO$ sample (with
grains of different shape and average size of the order of $1\mu m
$).

AC measurements were made by using a high-sensitivity homemade
susceptometer based on the screening method and operating in the
reflection configuration~\cite{9,14}. The complex response
$\chi_{ac}=\chi^{'}+i \chi^{''}$ was measured as a function of the
AC field $h_{ac}(t)=h \cos(\omega t)$ (applied normally to the
sample's surface with the amplitude $0.01Oe\le h\le 50Oe$ and
frequency $1kHz \le \omega \le 30kHz$) taken at fixed temperature.
The field dependence of the normalized real part of AC
susceptibility $\Delta \chi^{'}(h)=\chi^{'}(h)-\chi^{'}(0)$ for
different temperatures and $Pr$ content is shown in Fig.3. As it
is evident from this picture, there are two distinctive regions.
Namely, above $h=15Oe$ the curves exhibit almost linear dependence
which can be attributed to establishment of the
well-documented~\cite{18,19,20,21} Bean type critical state regime
with $1+4\pi  \chi_B^{'}(h)=2h/J_CD$ where $J_C$  is the
field-independent critical current density and $D$ the sample's
thickness. On the other hand, below $h=15Oe$, practically
temperature-independent periodic oscillations are clearly seen. In
what follows, we shall focus on explanation of this interesting
phenomenon.

\begin{figure}
\begin{center}
\includegraphics[width=6.20cm]{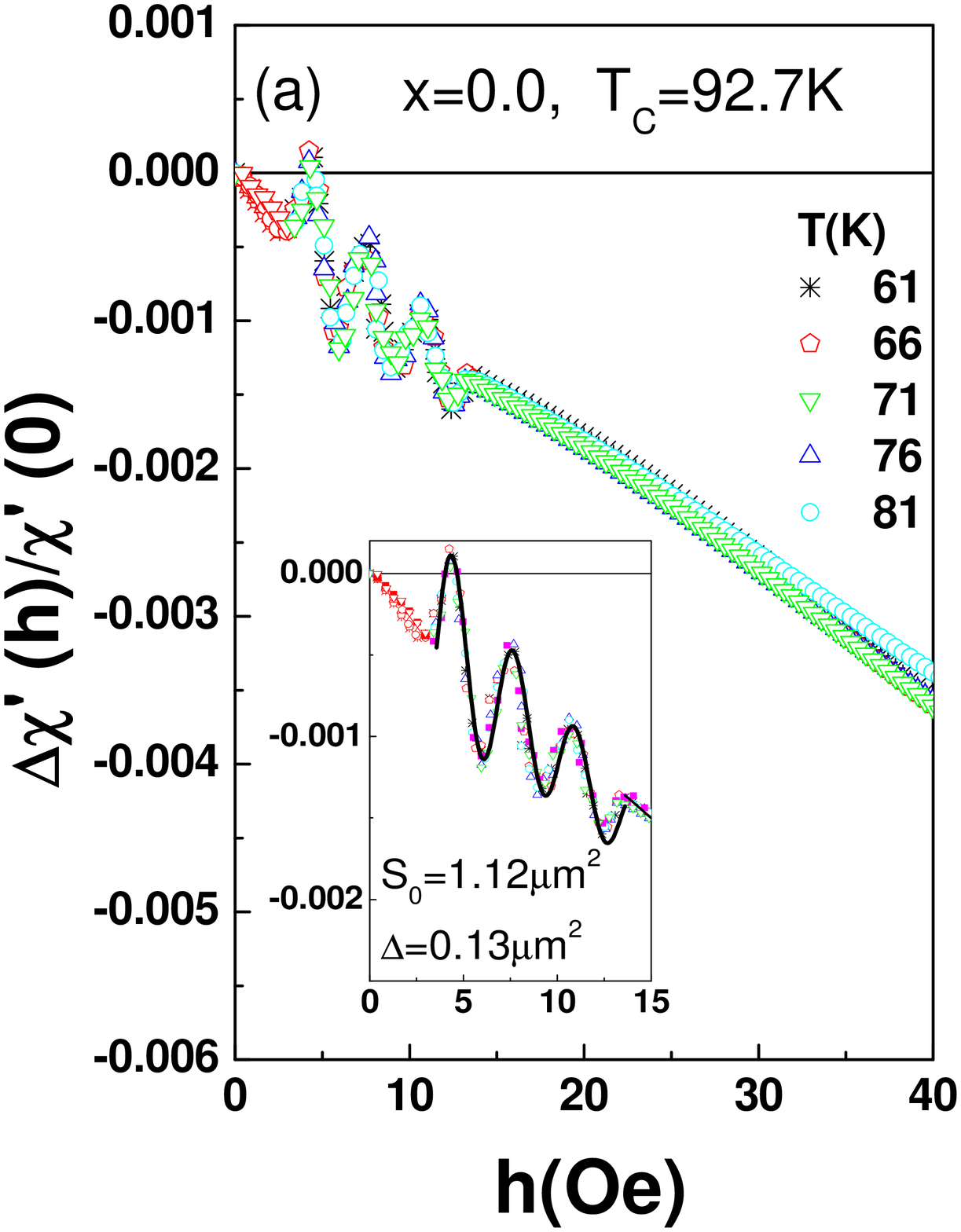}\vspace{0.05cm}
\includegraphics[width=6.20cm]{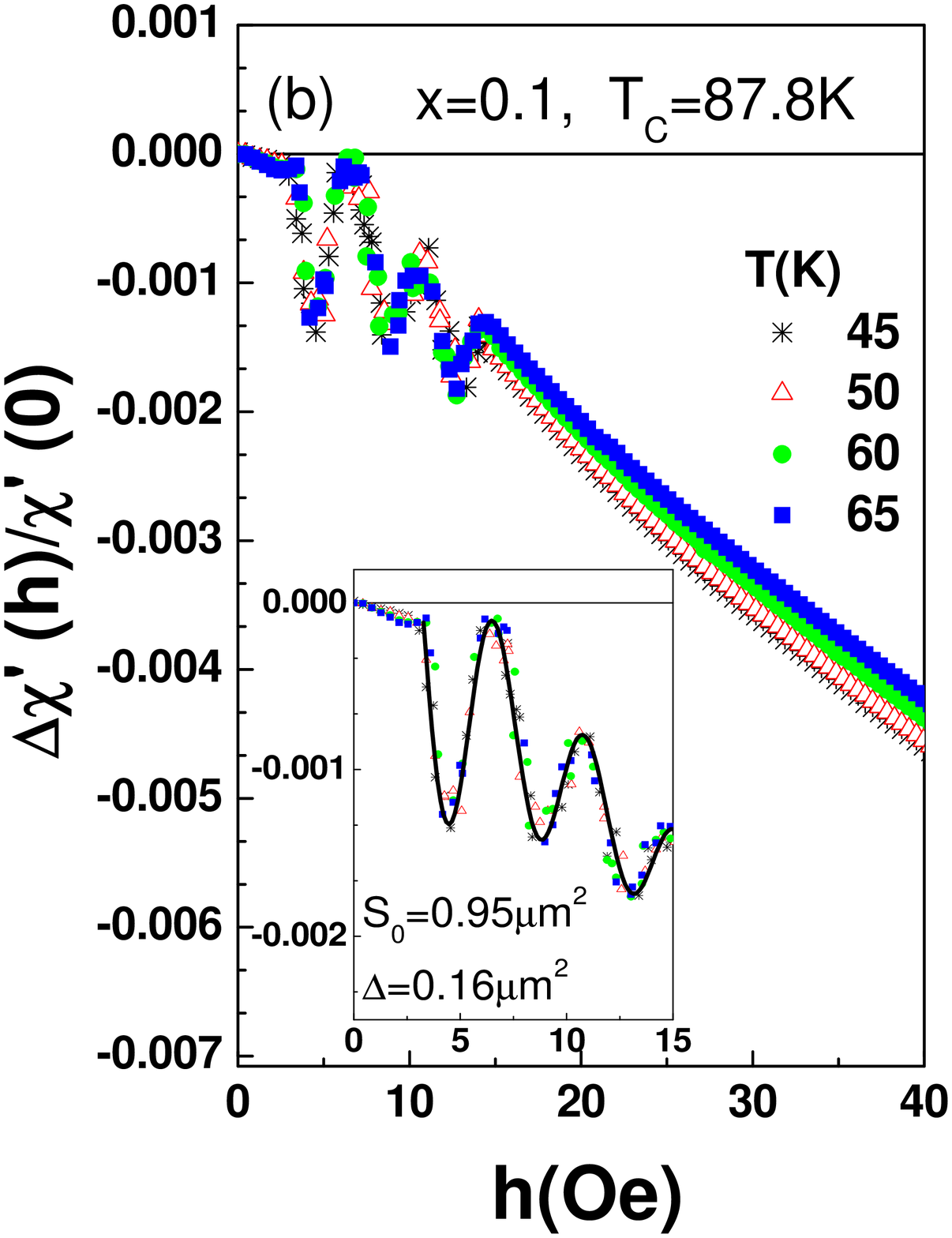}\vspace{0.05cm}
\includegraphics[width=6.20cm]{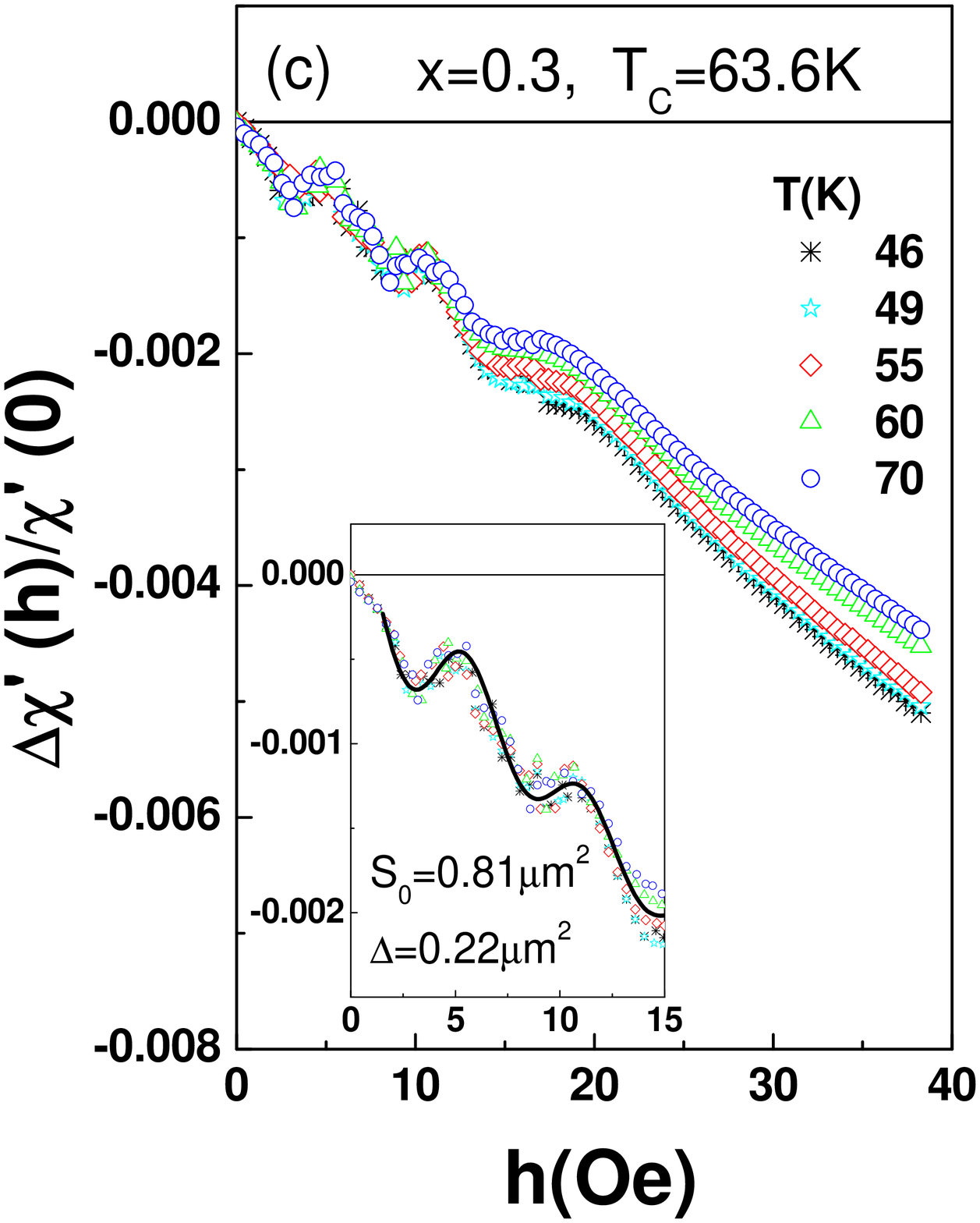}\vspace{0.0cm}
\caption{The magnetic field dependence of the normalized real part
of AC susceptibility at different temperatures for samples with
different $Pr$ content $x$. Inset: the best fits (solid line) of
the low-field region according to Eqs.(2)-(4).}
\end{center}
\end{figure}

\section{Discussion}

To understand the observed behavior of $\Delta \chi^{'}(h)$, it is
quite reasonable to assume that the low-field AC response in our
samples is related to a Josephson network mediated intergranular
contribution~\cite{22,23,24} $\chi_J^{'}(h)$. Notice, first of
all, that the field oscillations manifest themselves in a rather
narrow region between two characteristic Josephson fields: the
lower critical field $h_{c1}^J=\Phi_0/2\pi \lambda_J^2$ and the
upper critical field $h_{c2}^J=\Phi_0/2\pi \lambda_Jd$. Here,
$\lambda_J=\sqrt{\Phi_0/2\pi \mu_0J_Cd}$ is the Josephson
penetration depth (which also defines the size of a Josephson
vortex), and $d=2\lambda_L+l\simeq 2\lambda_L$ is the width of a
single contact with $\lambda_L$ being the London penetration depth
and $l$ the thickness of the insulating layer. Based on the above
critical fields, the flux penetration scenario can be described as
follows. For $h<h_{c1}^J$, we have analog of the Meissner state
for intergranular region (total screening of the applied field).
When $h>h_{c1}^J$, field starts to penetrate into the
intergranular region in the form of Josephson vortices (hence,
$h_{c1}^J$ is analog of the Abrikosov lower critical field
$h_{c1}^A=\Phi_0/2\pi \lambda_L^2$). Since (unlike Abrikosov
vortices) Josephson vortices are coreless, they will nucleate in
the contact area $S=Ld$ (where $L$ is the length of the contact)
until their distribution becomes homogeneous. This process takes
place for field region $h_{c2}^J>h>h_{c1}^J$ and manifests itself
in the Fraunhofer type dependence of the Josephson current
$I_J(h)=I_J(0)(\sin f/f)$ with $f=h/h_J$. The structure of the
pattern is governed by the characteristic Josephson field
$h_J=2\pi S/\Phi_0$ which indicates how many fluxons penetrate the
contact area $S$. At $h\simeq h_{c2}^J$, the Fraunhofer pattern
practically disappears. More precisely, $I_J(h_{c2}^J)\ll I_J(0)$
(hence, $h_{c2}^J$ indeed plays a role of the upper critical field
for Josephson vortices). For $h_{c1}^A>h>h_{c2}^J$, we have a
conventional Meissner state for the intragranular region (applied
field still can not penetrate inside grains). The formation of the
intragranular Abrikosov mixed state starts only for $h>h_{c1}^A$.

In order to clarify the origin of the discussed here effects, let
us estimate the values of the above critical fields (and the
corresponding depths). By relating the lower Josephson field
$h_{c1}^J=\Phi_0/2\pi \lambda_J^2$ with the beginning of the
observed oscillations (which is $h_{c1}^J\simeq 3.5Oe$ for pure
$YBCO$ sample), we obtain $\lambda_J\simeq 1\mu m$ for the size of
the Josephson vortex. On the other hand, by relating the upper
Josephson field $h_{c2}^J=\Phi_0/2\pi \lambda_Jd$ with the end of
oscillating behavior (which is $h_{c2}^J\simeq 15Oe$ for the same
sample), we get $\lambda_L\simeq 100nm$ as a reasonable estimate
of the London penetration depth in this material~\cite{16}. To
account for the observed evolution of the Josephson fields with
$Pr$ concentration $x$, we recall~\cite{17} that in addition to
the critical temperatures $T_C(x)\simeq T_C(0)(1-x)$, both the
London depth $\lambda_L(x)$ and the critical current density
$J_C(x)$ decrease upon doping. Namely, assuming that
$\lambda_L(x)\simeq \lambda_L(0)(1-x)$ and $J_C(x)\simeq
J_C(0)(1-x)$, we obtain $\lambda_J(x)\simeq \lambda_J(0)/(1-x)$,
$h_{c1}^J(x)=\Phi_0/2\pi \lambda_J^2(x)\simeq h_{c1}^J(0)(1-x)^2$
and $h_{c2}^J(x)=\Phi_0/4\pi \lambda_J(x)\lambda_L(x)\simeq
h_{c2}^J(0)$ for doping induced variation of the Josephson depth,
lower and upper Josephson fields, respectively. Notice first of
all that (in accord with our observations) the upper field does
not change with $Pr$ and has the same value ($h_{c2}^J\simeq
15Oe$) for all three samples. As for the lower field, according to
the above simplified expressions, we have $h_{c1}^J(x=0)\simeq
3.5Oe$, $h_{c1}^J(x=0.1)\simeq 0.8h_{c1}^J(x=0)\simeq 2.8Oe$, and
$h_{c1}^J(x=0.3)\simeq 0.49h_{c1}^J(x=0)\simeq 1.75Oe$ for $x=0$,
$x=0.1$, and $x=0.3$, respectively. All these estimates are in
good agreement with the observed onset of field oscillations (see
Fig.3). Besides, the above $x$-dependence of the London
penetration depth results in the following evolution of the
contact area $S(x)\simeq S(0)(1-x)$ where $S(0)\simeq
2L\lambda_L(0)$. Now we can also estimate the value of this area
for each of three samples by relating the number of trapped
fluxons $n(x)=2\pi hS(x)/\Phi_0$ with the number of the observed
oscillation minima (seen in Fig.3) for the applied field span
(lying between the lower and upper critical fields) $h\simeq
h_{c2}^J(x)-h_{c1}^J(x)$. Namely, using $n(0)=4$, $n(0.1)=3$, and
$n(0.3)=3$, we obtain $S(0)\simeq 1\mu m^2$, $S(0.1)\simeq 0.9\mu
m^2$, and $S(0.3)\simeq 0.7\mu m^2$ for the estimates of contact
areas in our three samples, which remarkably correlate with the
above assumed doping dependence of $S(x)$. Moreover, the doping
dependence of the London penetration depth $\lambda_L(x)$ also
controls the evolution of the lower Abrikosov field,
$h_{c1}^A(x)=h_{c1}^A(0)/(1-x)^2$ leading to the following
estimates: $h_{c1}^A(0)\simeq 0.03T$, $h_{c1}^A(0.1)\simeq 0.04T$,
and $h_{c1}^A(0.3)\simeq 0.06T$. Therefore, given a markedly
different values for Josephson and Abrikosov critical fields, we
can safely assume that the discussed here phenomenon is strictly
related to the Josephson physics. A pronounced Fraunhofer type
form of the observed curves suggests a rather strong coherent
response from many Josephson junctions comprising the
grain-boundary network (despite some distribution in sizes of the
individual junctions seen in Fig.2). To describe the observed
phenomenon, we assume that intergranular contribution
$\chi_J^{'}(h)$ is related to AC field $h_{ac}(t)=h \cos(\omega
t)$ induced modulation of the Josephson current
$I_{ij}(t)=(2\pi/\Phi_0)J_{ij}\sin \theta_{ij}(t)$ (where $J_{ij}$
is the Josephson energy) circulating in a closed plaquette
(cluster) with a random distribution over contact areas $S_{ij}$.
Each such cluster involves adjacent superconducting grains
$i=1,2,...N$ and $j=i+1$ with an effective phase difference
$\theta_{ij}(t)=2\pi S_{ij}h_{ac}(t)/\Phi_0$ across intergranular
barriers~\cite{9,23,24,25,26}. In turn, due to the Ampere's law,
this circulating current $I_{ij}(t)$ produces a net magnetic
moment~\cite{25,26} $\mu(t)=\sum_{ij}I_{ij}(t)S_{ij}$, leading us
to ${\cal H}(t)=\sum_{ij}{\cal H}_{ij}(t)$ for the total
Hamiltonian describing the flux dynamics of a single plaquette
with
\begin{equation}
{\cal H}_{ij}(t)=J_{ij}[1-\cos \theta_{ij} (t)]-
\frac{2\pi}{\Phi_0}J_{ij}\sin \theta_{ij}(t)h_{ac}(t)S_{ij}
\end{equation}
where the second term is a Zeeman contribution $\mu(t)h_{ac}(t)$.

To obtain the experimentally observed intergranular contribution
to the AC response, we assume (for simplicity) a Lorentz type
distribution of the contact areas $S_{ij}$ (around their mean
values $S_0$ with the width $\Delta$) of the form:
\begin{equation}
F(S_{ij})=\frac{1}{\pi}\frac{\Delta}{(S_{ij}-S_0)^2+\Delta^2}
\end{equation}
keeping in mind that the Josephson energy $J_{ij}$ also depends on
the contact area $S_{ij}$ via a distance between grains $r_{ij}$.
Namely, according to the conventional description of granular
superconductors, $J_{ij}=J(0)\exp(-r_{ij}/2r_0)$ where $r_0$ is of
the order of an average grain size (radius). By using some
geometrical arguments, it can be demonstrated that $S_{ij}\simeq
(r_{ij}/r_0)^2S_0$ which results in the following explicit
dependence of the Josephson energy on contact area,
$J_{ij}=J(0)\exp(-\sqrt{S_{ij}/4S_0})$. Notice that this way we do
not introduce any new fitting parameters (apart from the above
mentioned $S_0$ and $\Delta$).

Thus, the expression for the observed intergranular contribution
to the AC susceptibility finally reads
\begin{equation}
\chi^{'}(h)=<\chi_J^{'}(h)>_S=\sum_{ij}\int_{0}^{S_m}dS_{ij}F(S_{ij})\chi_{ij}^{'}(h)
\end{equation}
where
\begin{eqnarray}
\chi_{ij}^{'}(h)&=&\frac{1}{2\pi V}\int_{0}^{2\pi}d(\omega t)
\left[- \frac{\partial ^2 {\cal H}_{ij}(t)}{\partial h_{ac}^2(t)}
\right]\\ \nonumber &=&\chi _{ij}(0)\left
[J_0(f_{ij})-J_1(f_{ij})f_{ij} \right ]
\end{eqnarray}
Here, $\chi_{ij}(0)=(2\pi S_{ij}/\Phi_0)^2J_{ij}/V$ ($V$ is the
properly defined volume), $J_n$ are the Bessel functions, and
$f_{ij}=h/h_{ij}^J$, with $h_{ij}^J=\Phi_0/2\pi S_{ij}$ being a
characteristic Josephson field (which is eventually responsible
for the structure of the Fraunhofer pattern).

The best fits of the experimental data for the low-field region
based on Eqs.(2)-(4) along with the values of two fitting
parameters, $S_0$ and $\Delta$, are shown as insets in Fig.3 (we
assume $S_m=2S_0$). In particular, for undoped YBCO sample, we
found $S_0(0)=1.12\mu m^2$ for the mean value of the contact area
(which corresponds to the characteristic Josephson field
$h_J(0)=\Phi_0/2\pi S_0(0)\simeq 2.8Oe$) and $\Delta(0)=0.13\mu
m^2$ for the contact area width distribution. The contact areas in
the doped samples are found to be best fitted by the following set
of parameters: $S_0(0.1)=0.95\mu m^2$ (equivalent to
$h_J(0.1)=\Phi_0/2\pi S_0(0.1)\simeq 3.3Oe$), $\Delta(0.1)=0.16\mu
m^2$, $S_0(0.3)=0.81\mu m^2$ (equivalent to $h_J(0.3)=\Phi_0/2\pi
S_0(0.3)\simeq 3.9Oe$), and $\Delta(0.3)=0.22\mu m^2$. Notice that
decreasing of $S_0(x)$ with $x$ closely follows the earlier
suggested doping dependence $S_0(x)\simeq S_0(0)(1-x)$, while the
opposite behavior of the widths $\Delta(x)$ (increasing in doped
samples) most likely reflects a random accumulation of $Pr$ on
grain boundaries, leading to a more broad distribution of the
contact areas.

In summary, by using a highly sensitive homemade AC magnetic
susceptibility technique, the magnetic flux penetration has been
measured in high-quality $Pr_xY_{1-x}Ba_2Cu_3O_{7-\delta}$
polycrystals as a function of AC magnetic field for different
temperatures and $Pr$ doping. In addition to the conventional
critical state behavior at higher fields, a clear manifestation of
coherent intergranular response from Josephson vortices seen as a
periodic Fraunhofer type dependence of the real part of AC
susceptibility was observed at low magnetic fields.

The authors gratefully acknowledge Brazilian agencies CNPq, CAPES
and FAPESP for financial support.

\end{document}